\documentclass[aps,prl,preprint]{revtex4}

\usepackage{epsfig}

\draft

\begin{document}
\author{Qiaoli Yang}
\address{Department of Physics, Nanjing University, Nanjing 210093, P. R. China}

\title{The effect of strong interactions on quarks' electromagnetic vector vertex}

\bigskip

\begin{abstract}This paper studies the correction of electromagnetic vector
vertex of quarks due to the strong interactions.
\bigskip


\bigskip

\end{abstract}

\maketitle Since quarks carry both electric charges and color
charges, an interesting question will be how the strong interactions
disturbs quarks' electromagnetic vector vertex. For electrons, the
tree-level electromagnetic vector function is a simple $\gamma_\mu$,
thus the corresponding "Dirac" magnetic moment, which is usually
expressed in term of the Lande g-factor, is: $2$. However, higher
order Feynman diagrams will differ the e-m vector vertex from a pure
$\gamma_\mu$, then g will slightly differ from 2. This is well known
as the "anomalous magnetic moment" of electron. Composite particles
often have a much huger anomalous magnetic moment such as proton.
This indicates a great difference of e-m vertex function from simple
$\gamma_\mu$.

In ordinary state free quarks do not exist, however it may exist in
strange matter under special environments such as in neutron stars.
So, it will useful to consider the "anomalous magnetic moment" of
quarks. This is essentially to calculate the e-m vertex function of
quarks with both higher order e-m feynman diagrams and effects of
the strong interactions. The higher order e-m feynman diagrams are
always considered as small perturbation, so we can safely
concentrate to latter contribution. In this paper I calculate the
e-m vertex function of free quark with a relatively simple model
"GCM" to describe the strong interactions, and employ
Dyson-schwinger equation as a non-perturbation method for strong
interactions calculation. (the Euclidean space field formulation is
also employed for calculation convenience in future studies; The
question of how one may proceed between Euclidean and Minkowski
space is discussed by[1].)

The GCM generating functional of massless quarks in Euclidean space
can be written as
\begin{equation}
Z[\bar{\eta},\eta, \emph{A}_{\mu}] = \int {\cal{D}}\bar{q}
{\cal{D}} q ~e^{\left\{-\int d^4 x~ \bar{q}(x)
[\gamma\cdot\partial +\gamma_{\nu}A_{\nu}(x)]q(x)-\int d^4 x d^4
y~\frac{1}{2}j^a_{\mu}(x)D_{\mu\nu}(x-y)j^a_{\nu}(y)+\int d^4
x~(\bar{\eta} q + \bar{q} \eta)\right\}}
\end{equation}
here  $j^a_{\mu}(x)$ denotes the color octet vector current $
j^a_{\mu}(x)=\bar{q}(x)\gamma_{\mu}\frac{\lambda^a_{C}}{2}q(x)$
and $D_{\mu\nu}(x-y)=\delta_{\mu\nu}D(x-y)$ denotes the gluon
two-point function.

The coordinate space vector vertex is:
\begin{equation}
\Gamma_{\mu}(x_1,x_2,y)=\frac{\delta^3\Gamma(A,\bar{q}^{cl},
q^{cl})}{\delta A_{\mu}(y)\delta\bar{q}^{cl}(x_1)\delta
q^{cl}(x_2)} \Big| _{{A_{\mu}}=0}
\end{equation}
Where $\Gamma(A,\bar{q}^{cl},q^{cl})$ denotes the effective
action. From the relation:
$$\frac{\delta^2\Gamma(A,\bar{q}^{cl},q^{cl})}{{\delta\bar{q}^{cl}(x_1)\delta{q}^{cl}(x_2)}}=g^{-1}[A_\mu]$$
Eq.(2) can be expressed as:
\begin{equation}
 \Gamma_{\mu}(x_1,x_2,
y)=\frac{\delta{g}^{-1}[A_\mu]}{\delta{A}_{\mu}(y)}\Big|_{{A_{\mu}}=0}
\end{equation}
here $g^{-1}[A_\mu]$ denotes the inverse quark propagator under
external field $A_{\mu}$ and can be expanded in powers of $A_\mu$
as follows:
\begin{eqnarray}
g^{-1}[A_\mu]&=&g^{-1}[A_\mu]\Big|_{{A_\mu}=0}+\int d^4 y
\frac{\delta g^{-1}[A_\mu]}{\delta A_\mu
(y)}\Big|_{A_\mu=0}A_{\mu}(y)+...\nonumber\\
&=&G^{-1}(x_1,x_2) +\int d^4y\Gamma_\mu(x_1,x_2,y)A_\mu(y)+...
\end{eqnarray}
Which leads to the formal expansion
\begin{equation}
g[A_\mu]=G-\int d^4y G A_\mu\Gamma_\mu G+...
\end{equation}

The dressed quark propagator has the decomposition in momentum
space:$$G^{-1}(p)=i\gamma\cdot pA(p^2)+B(p^2)$$where the functions
$A(p^2)$ and $B(p^2)$ are determined by the rainbow
Dyson-Schwinger equation.

In momentum space, Eq.(4), Eq.(5) lead to:$$G^{-1}\approx
g^{-1}[A_\mu ]-\Gamma_\mu A_\mu$$$$G_0^{-1}\approx
g_0^{-1}[A_\mu]-\Gamma^0_\mu A_\mu$$ and$$G\approx g[A_\mu]+GA_\mu
\Gamma_\mu G$$ Insert them to the rainbow equation
$$G^{-1}=G^{-1}_0+\frac{4}{3}\int\frac{d^4k}{(2\pi)^4}D_{\mu\nu}(p-k)\gamma_\nu G(k)\gamma_\mu$$
and up to the first order, one gets:
\begin{equation}
\Gamma_\mu(p,q)=\Gamma_\mu^0-\frac{4}{3}\int \frac{d^4k}{(2\pi)^4}
D(p-k)\gamma_\nu G(k+q/2)\Gamma_\mu(k,q)G(k-q/2)\gamma_\nu
\end{equation}

$\Gamma_\mu$ should be constrained by the vector Ward-Takahashi
identity:
\begin{equation}
q_\mu\Gamma_\mu(p,q)=G^{-1}(P+\frac{q}{2})-G^{-1}(p-\frac{q}{2})
\end{equation}

Since $\Gamma_\mu$ transforms as a vector, and contains parameters
$p$ and $q$, it can be written as[3]:
\begin{equation}
\Gamma_\mu(p,q)=\Lambda_1(p,q)\gamma_\mu+\Lambda_2(p,q)p_\mu+\Lambda_3(p,q)q_\mu
\end{equation}

Insert Eq.(8) to Eq.(7):
\begin{eqnarray}
q_\mu\Gamma_\mu=\Lambda_1q_\mu\gamma_\mu+\Lambda_2p\cdot
q+\Lambda_3q^2 & = & i\gamma\cdot(p+\frac{q}{2})
A\Big((p+\frac{q}{2})^2\Big)+B\Big((p+\frac{q}{2})^2\Big)
\nonumber\\&&
-\left\{i\gamma\cdot(p-\frac{q}{2})A\Big((p-\frac{q}{2})^2\Big)+B\Big((p-\frac{q}{2})^2\Big)\right\}
\end{eqnarray}
Take traces of Eq.(9) of both sides:
\begin{equation}
\Lambda_2 p\cdot q+\Lambda_3
q^2=B\Big((p+\frac{q}{2})^2\Big)-B\Big((p-\frac{q}{2})^2\Big)
\end{equation}

Dot $p_\mu$ on both sides of Eq.(6) one gets:
\begin{eqnarray}
&&p_\mu\Gamma_\mu = \Lambda_1
p_\mu\gamma_\mu+\Lambda_2p^2+\Lambda_3 p\cdot q \nonumber\\&& =
p_\mu\gamma_\mu -\frac{4}{3}p_\mu\int \frac{d^4k}{(2\pi)^4}
D(p-k)\gamma_\nu G_1\Gamma_\mu G_2\gamma_\nu
\end{eqnarray}

Dot $\gamma_\mu$ leads to:
\begin{eqnarray}
&&\gamma_\mu\Gamma_\mu = \Lambda_1
\gamma_\mu^2+\Lambda_2\gamma_\mu p_\mu +\Lambda_3 \gamma_\mu
q_\mu\nonumber\\&& = \gamma_\mu^2 -\frac{4}{3}\gamma_\mu\int
\frac{d^4k}{(2\pi)^4}D(p-k)\gamma_\nu G_1\Gamma_\mu G_2\gamma_\nu
\end{eqnarray}

To express above equations clearer I first evaluate expression
$$\gamma_\nu
G(k+\frac{q}{2})\Gamma_\mu(k,q)G(k-\frac{q}{2})\gamma_\nu$$ put
expression of $G(p)$ in it and do some calculations:
\begin{eqnarray}
&&\gamma_\nu G(p_1)\Gamma_\mu(k,q)G(p_2)\gamma_\nu\nonumber\\
&&=\gamma_\nu \frac{1}{i\gamma\cdot p_1 A(p_1^2)+B(p_1^2)}
[\Lambda_1\gamma_\mu+\Lambda_2k_\mu+\Lambda_3q_\mu]\frac{1}{i\gamma\cdot p_2A(p_2^2)+B(p_2^2)}\gamma_\nu \nonumber\\
&&=\frac{1}{[p_1^2A^2(p_1^2)+B^2(p_1^2)][p_2^2A^2(p_2^2)+B^2(p_2^2)]}\left\{I+II+III\right\}
\end{eqnarray}
Here $p_1=k+\frac{q}{2},p_2=k-\frac{q}{2}$ and:
\begin{eqnarray}
\lefteqn{I=\Lambda_1\{2A(p^2_1)A(p^2_2)(\gamma\cdot
p_2)\gamma_\mu(\gamma\cdot
p_1)-4iA(p^2_1)B(p^2_2)p_{1\mu}  }\nonumber\\
&&{} -4iA(p^2_2)B(p^2_1)p_{2\mu}-2B(p^2_1)B(p^2_2)\gamma_\mu\}
\end{eqnarray}
\begin{eqnarray}
\lefteqn{II=\Lambda_2k_\mu\{-4A(p^2_1)A(p^2_2)p_1\cdot
p_2+2iA(p^2_1)B(p^2_2)(\gamma\cdot p_1){}}\nonumber\\
&&{}+2iA(p^2_2)B(p^2_1)(\gamma\cdot p_2)+4B(p^2_1)B(p^2_2)) \}
\end{eqnarray}
\begin{eqnarray}
\lefteqn{III=\Lambda_3q_\mu\{-4A(p^2_1)A(p^2_2)p_1\cdot p_2
+2iA(p^2_1)B(p^2_2)(\gamma\cdot p_1){}\}}\nonumber\\
&&{}+2iA(p^2_2)B(p^2_1)(\gamma\cdot p_2)+4B(p^2_1)B(p^2_2))\}
\end{eqnarray}

In order to have a solution of $\Gamma_\mu$ function, insert
expression(13) to Eq.(11), Eq.(12) and take traces on both sides:
\begin{eqnarray}
4p^2\Lambda_2+4p\cdot q\Lambda_3&=&-\frac{4}{3}p_\mu\int
\frac{d^4k}{(2\pi)^4}D(p-k)tr
\{Eq.(13)\}{}\nonumber\\
&=&-\frac{4}{3}\int\frac{d^4k}{(2\pi)^4}D(p-k)
\frac{16}{[p_1^2A^2(p_1^2)+B^2(p_1^2)][p_2^2A^2(p_2^2)+B^2(p_2^2)]}\nonumber\\
&&{}\ast\Big\{\left[-iA(p^2_1)B(p_2^2)p\cdot p_1-iA(p^2_2)B(p^2_1)p\cdot p_2\right]\Lambda_1 \nonumber\\
&&{}+p\cdot k\left[-A(p^2_1)A(p^2_2)p_1\cdot
p_2+B(p^2_1)B(p^2_2)\right]\Lambda_2\nonumber\\
&&{}+p\cdot q\left[-A(p^2_1)A(p^2_2)p_1\cdot
p_2+B(p^2_1)B(p^2_2)\right]\Lambda_3\Big\}
\end{eqnarray}
and
{\setlength\arraycolsep{2pt}
\begin{eqnarray}
16\Lambda_1&=&16-\frac{4}{3}\int\frac{d^4k}{(2\pi)^4}D(p-k){}
tr\{\gamma_\mu Eq.(13)\} {} \nonumber\\
&&=16-\frac{4}{3}\int\frac{d^4k}{(2\pi)^4}D(p-k)
\frac{1}{[p_1^2A^2(p_1^2)+B^2(p_1^2)][p_2^2A^2(p_2^2)+B^2(p_2^2)]}\nonumber\\
&&\ast\Big\{-16[A(p^2_1)A(p^2_2)p_1\cdot
p_2+2B(p^2_1)B(p^2_2)]\Lambda_1\nonumber\\
&&+[(8iA(p^2_1)B(p^2_2)k\cdot p_1+8iA(p^2_2)B(p^2_1)k\cdot
p_2)]\Lambda_2
\nonumber\\
&&+[(8iA(p^2_1)B(p^2_2)k\cdot p_1+8iA(p^2_2)B(p^2_1)k\cdot
p_2)]\Lambda_3\Big\}
\end{eqnarray}}
These two equations above combined with Eq.(10) and the rainbow
Dyson-Schwing equation of functions $A(p^2)$ and $B(p^2)$
\begin{eqnarray}
&&[A(p^2)-1]p^2=\frac{8}{3}\int\frac{d^4q}{(2\pi)^4}D(p-q)\frac{A(q^2)p\cdot
q}{q^2A^2(q^2)+B^2(q^2)}\nonumber\\
&&B(p^2)=\frac{16}{3}\int\frac{d^4q}{(2\pi)^4}D(p-q)\frac{B(q^2)
}{q^2A^2(q^2)+B^2(q^2)}
\end{eqnarray}
will in principle fully restrict the nonperturbative vector vertex
of quarks. By numerically study Eq.(10) Eq.(17) Eq.(18) and
Eq.(19) with appropriate gluon two-point function, one can get the
nonperturbative vector vertex and quark propagater in Euclidean
space.

In order to have a qualitative understanding of this method, a
particularly simple model gluon two-point function is used as
follows:
\begin{equation}
D\Big(p-k\Big)=g\delta^4(p-k)
\end{equation}
when $q\ll p$, we have $p+q/2\approx p-q/2\approx p$, Eq(10),
Eq(17), Eq(18) have a simple solution of $\Lambda_i$ with
parameters $A(p^2)~and~B(p^2)$  :
\begin{eqnarray}
&&\Lambda_1=\frac{12\pi^4[B^2+3\pi^4(A^2p^2+B^2)^2/g-A^2p^2]}{g+36\pi^8(A^2p^2+B^2)^2/g-15\pi^4A^2p^2+6\pi^4B^2}\nonumber\\
&&\Lambda_2=\frac{24iAB\pi^4}{g+36\pi^8(A^2p^2+B^2)^2/g-15\pi^4A^2p^2+6\pi^4B^2}\nonumber\\
&&\Lambda_3=\frac{24iAB\pi^4p\cdot q}{\Big[g+36\pi^8(A^2p^2+B^2)^2/g-15\pi^4A^2p^2+6\pi^4B^2\Big]q^2}\nonumber\\
\end{eqnarray}
Here A, B are solutions of the rainbow equation(19)

Insert the same two-point function of gluon to Eq.(19):
\begin{equation}
B(p^2)=0,~~~A(p^2)=\frac{1}{2}\Big[1-\Big(1+\frac{2g}{3\pi^4p^2}\Big)^{1/2}\Big]
\end{equation}
\begin{equation}
B(p^2)=0,~~~A(p^2)=\frac{1}{2}\Big[1+\Big(1+\frac{2g}{3\pi^4p^2}\Big)^{1/2}\Big]
\end{equation}
and:
\begin{equation}
B(p^2)=(\frac{g}{3\pi^4}-4p^2)^{1/2},~A(p^2)=2 ~~for~~
p^2\le\frac{g}{12\pi^4}
\end{equation}

Solution(24) generates a dynamical quark mass and breaks chiral
symmetry spontaneously. Insert this solution to Eq.(21), I get the
e-m vertex function of quarks:
\begin{equation}
\Gamma_\mu=\frac{8}{7}\gamma_\mu+\frac{16i}{7(\frac{g}{3\pi^4}-4p^2)^{1/2}}p_\mu+\frac{16ip\cdot
q}{7(\frac{g}{3\pi^4}-4p^2)^{1/2}q^2}q_{\mu}
\end{equation}

\bigskip

\vspace*{0.2 cm} \noindent{\large \bf Acknowledgments}

I would like to express my deep thanks to people who have ever
helped me, especially to professor Hong-shi Zong and professor Fan
Wang.

\bigskip

\vspace*{0.2 cm} \noindent{\large \bf References}
\begin{description}
\item{[1]} Craig D. Roberts and Anthony G. Williams, arXiv:
hep-ph/9403224 11 Aug 1997. \item{[2]}Hong-shi Zong, Xiang-song
Chen, Fan Wang, Chao-Hsi Chang and En-guang zhao, Phys. Rev. C66,
015201(2002). \item{[3]} Michael E. Peskin, Daniel V. Schroeder,
An Introduction to Quantum Field Theory, page 186.
\end{description}

\end{document}